\documentstyle[12pt]{article}    

\begin{document}

\title{ Production of kinks in an inhomogenous medium. }

\author{
{\sc T. Dobrowolski}
\\ Institute of Physics AP,
\\ Podchor\c{a}\.zych  2, 30-084 Cracow, Poland \\
e-mail: sfdobrow@cyf-kr.edu.pl}

\maketitle

\begin{abstract}
The purpose of this report is presentation of the main
modifications of the standard Kibble-Zurek formalism caused by the
existence of  unperfections in the system. We know that the
distribution  of kinks created during a second order phase
transition in pure systems is determined solely by the correlation
length at freeze-out time. The correlation length at that instant
of time intuitively describes the size of the defect and therefore
the number density of defects is limited by the possibility of
holding kinks in a unit volume. On the other hand if the system is
populated by the impurities then kinks emerge mainly in knots of
the force distribution which correspond to extremes of the
impurity potential i.e. positions of imperfections. The purpose of
this report is to show that, due to existence of the strong
gradients of the impurity potential, kinks can be created mainly
in the close vicinity of the impurities. It seems that this simple
mechanism can be responsible for occurrence, in the number density
formula, the additional length scale describing the impurity
distribution. We know that in pure systems, as a consequence of
kink-antikink annihilation, the number density of kinks decrease
in time. In contradiction to pure systems, kinks produced in the
systems populated by impurities could be confined by the impurity
centers and therefore they may not disappear from the system and
may remain above the level established by thermal nucleation of
pairs.
\end{abstract}

\section{Introduction}
Last years topological defects attract attention of many
researchers. The motivation of these studies comes from the fact
that they can be seen as macroscopic manifestations of underlying
physical processes. On the other hand they can help to study the
nature of critical dynamics.

The  theory describing the dynamics of the second order phase
transition was proposed by Kibble and Zurek \cite{Kibble}. The key
point of the Kibble-Zurek mechanism is an observation that the
order parameter evolves adiabatically through a sequence of nearly
equilibrium configurations up to the freeze-in time. At that
instant the system loses capacity to respond for the changes of
the external parameters. From that time to the freeze-out time the
field configuration remains almost unchanged. The dynamical
evolution restarts below the critical temperature at freeze-out
time. At that instant the system regains ability to respond for
the changes of the external parameters but it is too late to undo
non-trivial arrangements of the order parameter from above the
critical point. This paradigm works for the overdamped and
underdamped systems as well. The main prediction of this scenario
is the dependence of the number density of produced defects on
correlation length $n \sim \xi^{-d}$ at freeze-out time or its
dependence on quench time $n \sim \tau^{- d/4}$, where $d$ denotes
the number of space dimensions. This scenario was well verified in
a series of numerical experiments \cite{Bray}.

The defect network
density obtained at freeze-out time is an initial condition for
dynamics which is determined by the defect - antidefect
interactions. Due to annihilation of defects and antidefects the
initial density of defect network first is quickly reduced in time and
then is stabilized on the level determined by the Boltzman factor which
describes the probability of thermal nucleation of the kink-antikink
pairs \cite{Buttiker}.

In real life experiments researchers use mainly stable coherent
quantum systems. So far,  experiments were performed in Helium-3
on symmetric phase $^3$He-B which is more simple to experimental
and theoretical treatment. The results of experimental studies
confirms the dependence of the number density of produced vortices
on quench time \cite{Ruutu}. Importance of studies of the
transitions in Helium-3, follows from the fact that due to
nontrivial structure of the order parameter it allows for
experimental verifications of ideas concerning the structure of
vacuum of the quantum field theory.

Researchers performed also experiments on liquid crystals. The
creation of disclinations of different types produced during a
quench from disordered to nematic phase in liquid crystals was
examined and the results were to some degree consistent with the
Kibble-Zurek predictions \cite{Chandar}.

There were also more controversial experiments made in superfluid
helium-4 where almost no vortices of
topological origin were observed \cite{Dodd}.

Lately there are also attempts to
study the creation of vortices in optically cooled alkali atom
clouds during formation of the Bose-Einsten condensate.

In this report I would like to concentrate on influence of impurities on
creation of topological defects. It is difficult to imagine free of
unperfections liquid crystal  or even superconductor.
The population of the superconductors and liquid crystals by the impurities and
admixtures seems to be an inevitable outcome of their preparation.
On the other hand quantum liquids are one of the purest  substances in the nature.
Although the solubility of foreign materials in liquid helium is almost zero
there exists some artificial techniques, like aerogel technique \cite{Lawes},
which  allow to  introduce impurities even into quantum liquid.

Prevailed part of the obtained hitherto results concern homogenous medium.
On the other hand, the presence of the impurities can significantly
change properties of the system.

This report aims in presentation of the main modifications of the
standard Kibble-Zurek formalism caused by the existence of
unperfections in the system.

\section{An influence of inhomogenities on production of kinks}

First let us recall K-Z formalism applied to description of
homogenous systems. A pure, overdamped $\phi^4$ system is
described by the following equation of motion
\begin{equation}
\gamma \partial_t \phi(t,x) = \partial^2_x \phi(t,x) - a(t)
\phi(t,x) - \lambda  \phi^3(t,x) + \eta(t,x),
\end{equation}
where  $\eta(t,x)$ is a temperature white gaussian noise defined
by the correlators
$$
\langle {\eta}(t,x) \rangle = 0,
$$
\begin{equation}
\langle {\eta}(t, x) {\eta} (t', x') \rangle = {2 \pi \gamma \over
\beta} \delta(x - x') \delta(t-t') .
\end{equation}
The explicit time dependence of the "mass" parameter $a(t)$ allows
for modelling the phase transition in the system. Depending on the
sign of this parameter we can find a system in the phase with one
trivial or two nontrivial ground states.

The number of kinks produced during a phase transition at freeze-out time
is calculated from the Liu-Mazenko-Halperin formula \cite{Liu}
\begin{equation}
n = {1 \over \pi} \sqrt{\langle \phi'^2 \rangle  \over \langle
\phi^2 \rangle }.
\end{equation}
In fact this formula can be expressed with the use of the  power
spectrum which is defined by the equal time correlator of the
order parameter.  The cut-off in this formula separates the stable
from unstable modes of the system. We integrate only over unstable
modes because only they can grow to form stable kink structures.
As a result of this calculation one could obtain a Kibble-Zurek
critical exponent $1/4$ which describes the dependence of the
number density of produced  kinks on quench time $\tau$
\begin{equation}
n \sim {1 \over \tau^{1 \over4 }}.
\end{equation}
In this context usually a linear quench is presumed.  The critical
exponent depends only on number of spatial dimensions.

On the other hand, many physical systems are dense populated by
imperfections of different types. The presence of impurities in
the system can be taken into account by introducing a
deterministic force distribution ${\cal D}(t,x)$ into the equation
of motion
\begin{equation}
\gamma \partial_t \phi(t,x) = \partial^2_x \phi(t,x) - a(t)
\phi(t,x) - \lambda  \phi^3(t,x) + \eta(t,x) + {\cal D}(t,x).
\end{equation}

First we reconsider the way of counting zeros of the order
parameter. The number density of produced zeros can be defined as
a ratio of zeros located in some interval of space to the length
of this interval. On the other hand the number of  zeros can be
calculated as a sum of arbitrary quantity divided by itself  over
all points where the scalar field disappears.  One of the possible
choices of this quantity is $\phi'$
\begin{equation}
n(t,x) = \lim_{L \rightarrow 0} \frac{\langle N \rangle}{2 L} =
\lim_{L \rightarrow 0} \frac{1}{2 L}\langle \sum_i
\frac{|\phi'(t,x_i)|}{|\phi'(t,x_i)|}\rangle .
\end{equation}
We know that integration of the delta function can be replaced by the
summation over zeros of delta argument
$$
\int d x f(x) \delta[g(x)]= \sum_i \frac{f(x_i)}{|g'(x_i)|}\;
\hbox{where $x_i$ are defined by the equation}\; g(x_i)=0.
$$
If we identify in this lemma function  $f$  with  $\phi'$  and  $g$ with  $\phi$ then
we replace the sum over all zeros by the average of some integral
\begin{equation}
n(x) = \lim_{L \rightarrow 0} \frac{1}{2 L}\langle
\int_{x-L}^{x+L} d \tilde{x} |\phi'(t,\tilde{x})|
\delta[\phi(t,\tilde{x})]\rangle .
\end{equation}
In zero $L$ limit this expression reduces  to the average over all realizations
of the noise of some combination of the delta and sign functions
\begin{equation}
n(x)= \langle sign[\phi'(t,x)] \phi'(t,x) \delta[\phi(t,x)]\rangle .
\end{equation}
In the next step we use integral representations of the signum and
delta functions and then divide the scalar field on two components
$\phi(t,x) = \psi(t,x) + u(t,x)$. One which carries deterministic
part of the evolution $u$ and the second $\psi$ which carries the
stochastic part of the  evolution of the scalar field $\phi$. We
known that above critical point  the field fluctuates around
trivial ground state. If the amplitude of the noise is small i.e.
if the temperature of the system is low then fluctuations of the
field are also small and  its value is close to zero. As the field
fluctuate around zero value its average magnitude is small and
therefore cubic term in the equation of motion is negligible. As
the system evolves through a sequence of almost equilibrium states
this remains true up to freeze-in time. Identification of those
two components in linear approximation is straightforward.

Then we use theorems which allow to replace $n$-th order
correlators by the average and correlator of second order and
therefore we are able to replace the averages of some functions of
random variable  by the functions of the second order correlator
of this variable.

Finally we obtain a formula which in the absence of impurities
reduces to the well known Halperin-Liu-Mazenko formula (3)
\begin{equation}
n(t,x) = {1 \over \pi} \sqrt{\langle \psi'^2 \rangle \over \langle
\psi^2 \rangle} e^{- {u^2 \over 2 \langle \psi^2 \rangle} - {u'^2
\over 2 \langle \psi'^2 \rangle}} + {u' \over \sqrt{ 2 \pi \langle
\psi^2 \rangle}} e^{- {u^2 \over 2 \langle \psi^2 \rangle}}
Erf{\left({u' \over \sqrt{2 \langle \psi'^2 \rangle}} \right)},\end{equation}
where $Erf$ is the error function. The analytical and numerical
studies of the kink distribution shows that kinks are created
mainly in the vicinity of knots of the force distribution which
corresponds to extremes of the impurity potential
\cite{Dobrowolski1}.

In most of the physical systems the shape and the distribution of the impurities
is random and therefore we have to allow the force distribution to be a random type
with some length scale which characterizes  the average distance between impurities.
This time the equation of motion contains two random forces. First represents thermal
fluctuations in the system $\eta$ and the second one which describes distribution
of impurities ${\cal D}$.

The angle bracket represents the average with respect to all realizations of the
thermal noise and the new bracket $\{...\}$ represents an average with respect
to all possible distributions of the  impurities in the system.

Equation of motion in this new situation, at first sight, seems to
be identical with equation (5)
\begin{equation}
\hat{\gamma} \partial_t \phi(t,x) = \partial^2_x \phi(t,x) - a(t)
\phi(t,x) - \lambda  \phi^3(t,x) + \eta(t,x) + {\cal D}(t,x).
\end{equation}
In fact $\hat{\gamma}$ in this equation is not a simple constant but it is an integral
operator. The existence of this term is an inevitable if we
restrict our studies to stationary processes. The explicit form of
this operator is the following:
$$
\hat{\gamma} \partial_t \phi(t,x) \equiv \int_{t_0}^t d t' \int_{-
\infty}^{\infty} d^3 x' \gamma(t,t';\vec{x}-\vec{x}')
\partial_{t'} \phi(t',\vec{x}').
$$
If the impurity force distribution ${\cal D}$ has a form of the
gaussian white noise then $\gamma(t,t';x-x') = \gamma \delta(t-t')
\delta(x-x')$, and this integral reduces to the damping constant
multiplied by the time derivative of the order parameter. In
generic situation we expect a dependence of ${\cal D}$ correlators
on some length scale (e.g. average distance between impurity
centers) and therefore the force distribution is defined as
follows
$$
\textbf{\{} {\cal D}(t,{x}) \textbf{\}} = 0,
$$
\begin{equation}
\textbf{\{} {\cal D}(t, x) {\cal D} (t', x') \textbf{\}} = {1
\over \beta} W(|x-x'|)  \delta(t-t') .
\end{equation}

To better understand the source of this complication let us consider a simple
mechanical analogy which is a Brownian motion theory.

The erratic motion of a Brownian particle is caused by collisions
with the molecules of the fluid in which it moves. These
collisions allow an exchange of the energy between the fluid and
the Brownian particle. If the Brownian particle is much more
massive than the molecules of the fluid then the influence of the
molecules on the observed particle can be approximated by a  Gaussian white noise
$\eta_G(t) $:
\begin{equation}
m \ddot{x}(t) + \gamma \dot{x}(t) = \eta_G(t) ,
\end{equation}
where $x(t)$ is the  position of the Brownian particle. The
generalization of  Brownian motion theory to the random motion of
a particle which is not necessarily heavier than the molecules of
the fluid was proposed by Kubo \cite{Kubo}. In this case the time
scale of molecular motion is no longer very much shorter than that
of the motion of the particle under observation, so that the
random force $\eta(t)$ can not be of Gaussian type. To describe an
influence of the molecules on the observed particle we have to
introduce a color noise characterized by some time scale. This
time scale may describe the average time interval between two
subsequent collisions of the molecules with the observed particle.
In addition, if we consider a stationary process we have to
abandon the assumption of a constant friction and to introduce
generally a frequency-dependent friction
\begin{equation}
m \ddot{x}(t) + \int^t_{t_0} d t' \gamma(t - t') \dot{x}(t') =
\eta(t) .
\end{equation}
In case of $\phi^4$ model the distribution of impurities is not
generally described by the white gaussian noise. In generic
situation the distribution of impurities is characterized by some
length scale which describes the average separation o impurity
centers and therefore, in similar way as it was in case of
Brownian particle, we introduce retardation to $\phi^4$ model.

Next we have to make further generalization of the
Liu-Mazenko-Halperin formula. This generalization is achieved by
averaging the formula (9) with respect to possible distributions
of the impurity centers
\begin{equation}
n = \{ n(t,x)\} = {1 \over \pi} \sqrt{\langle \psi'^2 \rangle +
\textbf{\{} u'^2 \textbf{\}} \over \langle \psi^2 \rangle +
\textbf{\{} u^2 \textbf{\}}}.
\end{equation}
The further understanding can be made for particular choice of the
noise amplitude. The most representative one is Ornstein-Uhlenbeck
amplitude $W(|x|) = {\cal A} e^{- \frac{|x|}{L}}$ which
interpolates between constant distribution and gaussian white
noise. In this model the number density of produced kinks depends
on quench time and on characteristic length scale of the impurity
distribution as well  \cite{Dobrowolski2}
\begin{equation}
n \approx \frac{1}{\pi} \sqrt{ \frac{0.43
  \frac{b}{\tau^{\frac{1}{2}}}  + 0.34
\frac{c}{ L^2 }  } {1.81 b  + 0.83 c }},
\end{equation}
where  $b= \frac{1}{\tau^{\frac{1}{4}}}  $ and $c= \frac{
L}{\sqrt{\tau}}$. Let us notice that in case of week imperfections the usual
scaling is recovered
\begin{equation}
n \sim {1 \over \tau^{1 \over 4 }}.
\end{equation}

\section{Remarks}
We show that there are two components which determine the number
density of kinks produced in the inhomogenous system during the
second order phase transition. First component follows from the
K-Z formalism for pure systems and is determined by the quench
time. The second component is determined by the characteristic
length which describes the distribution of impurities in the
system. Due to existence of impurities and admixtures the kinks
are created mainly in the knots of the impurity force
distribution. One could even find exact solutions which describe
the kinks confined by some particular impurity potentials
\cite{Dobrowolski3}. An examples of those solutions are squeezed
kink or squeezed anti-kink. One could also check linear stability
of those solutions. The other solution obtained for some
particular force distribution is a static kink-antikink solution
which illustrates that, in contradiction to pure systems, the
configuration of this type have no tendency to annihilate. This is
also a reason why at late time after transition in the
inhomogenous system we still can  find the number density of kinks
substantially larger than estimated from the probability of
thermal nucleation of kink-antikink pairs.

\hfill\break {\bf Acknowledgements}\hfill\break I am indebted to
organizers of the "Third COSLAB and First Joint COSLAB - VORTEX -
BEC2000+" workshop for hospitality and also for supporting my stay
in Bilbao. Work supported in part by KBN grant 2 PO3B 025 25 and
ESF "COSLAB" Programe.

\end{document}